\documentclass{aa}  
\usepackage{graphicx}
\usepackage{txfonts}

\begin{document}
\title{Multi-wavelength observations\thanks{Based on observations carried 
out using the Advanced Camera for Surveys at the Hubble Space Telescope under 
Program ID 10496; the Very Large Telescope at the ESO Paranal Observatory under 
Program IDs 176.A-0589(A), 276.A-5034(A) and the New Technology Telescope at the 
ESO La Silla Observatory under Program ID 078.A-0265(B)} of a rich galaxy cluster 
at z $\thicksim$ 1: }
  \subtitle{the HST/ACS colour-magnitude diagram}

   \author{J. S. Santos, \inst{1,2}
          P. Rosati,\inst{3}
	  R. Gobat,\inst{3}
	  C. Lidman,\inst{4}
          K. Dawson,\inst{5}
	  S. Perlmutter,\inst{5}
          H. B\"ohringer,\inst{2}
	  I. Balestra,\inst{2} \\
	  C.R. Mullis, \inst{6} 
	  R. Fassbender,\inst{2}
	  J. Kohnert,\inst{7}  
	  G. Lamer,\inst{7}
 	  A. Rettura,\inst{8}
	  C. Rit\'e,\inst{3}
	  A. Schwope\inst{7}
	}

   \institute{\inst{1} INAF-Osservatorio Astronomico di Trieste,
              Via Tiepolo 11, 34131 Trieste, Italy\\
	     \inst{2} Max-Planck-Institut f\"ur extraterrestrische Physik,
              Giessenbachstra\ss e, 85748 Garching, Germany\\
             \email{jsantos@oats.inaf.it} \\
             \inst{3} European Southern Observatory, Karl Schwarzschild Strasse 2, Garching bei Muenchen, D-85748, Germany\\
             \inst{4} European Southern Observatory, Alonso de Cordova 3107, Casilla 19001, Santiago, Chile\\
             \inst{5} E.O. Lawrence Berkeley National Laboratory, 1 Cyclotron Rd., Berkeley, CA 94720\\
             \inst{6} Wachovia Corporation, NC6740, 100 N. Main Street, Winston-Salem, NC 27101 \\
             \inst{7} Astrophysikalisches Institut Potsdam (AIP), An der Sternwarte 16, D-14482 Potsdam, Germany \\
	     \inst{8} Department of Physics and Astronomy, Johns Hopkins University, Baltimore, MD21218, USA
             }

   \date{Received ... ; accepted ...}

 
  \abstract
   {XMMU J1229+0151 is a rich galaxy cluster with redshift $z$=0.975, that was serendipitously 
detected in X-rays within the scope of the XMM-Newton Distant Cluster Project. 
HST/ACS observations in the $i_{775}$ and $z_{850}$ passbands, as well as VLT/FORS2 spectroscopy 
were further obtained, in addition to follow-up Near-Infrared (NIR) imaging in J- and Ks-bands 
with NTT/SOFI.
 }
   { We investigate the photometric, structural and spectral properties of the early-type 
galaxies in the high-redshift cluster XMMU J1229+0151. }
   {Source detection and aperture photometry are performed in the optical and NIR imaging. 
Galaxy morphology is inspected visually and by means of Sersic profile fitting to the 21 
spectroscopically confirmed cluster members in the ACS field of view. 
The $i_{775}$-$z_{850}$ colour-magnitude relation (CMR) is derived with a method 
based on galaxy magnitudes obtained by fitting the surface brightness of the galaxies with Sersic models. 
Stellar masses and formation ages of the cluster galaxies are derived by fitting the observed 
spectral energy distributions (SED) with models based on Bruzual \& Charlot 2003.
Star formation histories of the early-type galaxies are constrained through the analysis of 
the stacked spectrophotometric data.
}
   {The structural Sersic index $n$ obtained with the model fitting is in agreement with 
the visual morphological classification of the confirmed members, indicating a clear 
predominance of elliptical galaxies (15/21). 
The $i_{775}$-$z_{850}$ colour-magnitude relation of the spectroscopic members shows a very 
tight red-sequence with a zero point of 0.86$\pm$0.04 mag and intrinsic scatter equal to 0.039 mag.
The CMR obtained with the galaxy models has similar parameters.
By fitting both the spectra and SED of the early-type population we obtain a star formation 
weighted age of 4.3 Gyr for a median mass of 7.4 $\times 10^{10}$ M$_{\sun}$. 
Instead of an unambiguous brightest cluster galaxy (BCG), we find three bright 
galaxies with a similar $z_{850}$ magnitude, which are, in addition, the most massive cluster members, 
with $\sim$2 $\times 10^{11}$ M$_{\sun}$.
Our results strengthen the current evidence for a lack of significant evolution of the scatter 
and slope of the red-sequence out to z $\thicksim$ 1.}
 {} 

   \authorrunning{J.S.Santos et al.}
   \titlerunning{Multi-wavelength observations of a rich galaxy cluster at z $\thicksim$ 1}

  \maketitle


\section{Introduction}

Distant (z$\sim$1) galaxy clusters are unique astrophysical laboratories particularly 
suited to witness and study galaxy formation and evolution. 

Detailed studies of the properties of galaxies in large samples of high-redshift clusters 
are required to distinguish the two main galaxy formation scenarios, which have been 
under discussion for more than 30 years. In the \textit{monolithic picture}
(\cite{eggen}; \cite{larson}), massive galaxies are 
expected to be formed early from a single progenitor. In contrast, the \textit{hierarchical scenario} 
(\cite{toomre}; \cite{white}) predicts that elliptical galaxies should form later, 
through mergers. The behavior of early-type galaxies (ETGs), which are found to comprise both the most 
massive $and$ oldest systems, is the main cause for this debate. Indeed, it is now established that 
the star formation histories of ellipticals are mass-dependent from both observational 
(\cite{thomas}, \cite{vanderwel}) and theoretical studies (e.g. \cite{delucia04}, \cite{menci08}), 
such that low mass galaxies have more extended star formation histories than massive ones. 
This implies that the less massive galaxies have a lower formation redshift than the more massive 
systems, whose star formation histories are predicted to peak at z$\sim$5 (\cite{delucia06}). 
This scenario is commonly referred to as ``$downsizing$'' (\cite{cowie}).
Supporting this picture, there is strong observational evidence for the bulk of the stars in massive 
ellipticals to be already formed at redshift $>2$ (\cite{vand05}, 
\cite{holden}).

The colour-magnitude relation (CMR, \cite{visv77}, \cite{visv78}) is a fundamental 
scaling law used to assess the evolution of galaxy populations. The CMR of 
local clusters shows the existence of a tight Red Sequence (RS, \cite{bower}, \cite{depropris}) 
(\cite{gladders}, {\cite{baldry}}) formed of massive red elliptical galaxies undergoing passive evolution, 
and the analysis of its main parameters (zero point, scatter and slope) provides a means to quantify the 
evolution of the galaxies properties with redshift. It remains, nevertheless, unclear to what 
degree the CMR is determined by age and metallicity effects. 

The study of high-$z$ samples of galaxies is also important to provide information for the modelling 
of physical processes in semi-analytical techniques. Semi-analytical modelling (SAM) employing 
AGN feedback to prevent the overproduction of blue galaxies have recently succeeded in predicting 
a large amount of massive 
old galaxies (\cite{delucia06}, \cite{bower06}, \cite{menci06}, \cite{croton}, \cite{somerville}), 
however, several issues 
remain yet to be solved, such as the incapability to reproduce quantitatively the colour-bimodality 
in the colour-magnitude diagram and the scatter of the red-sequence, which is overestimated by a 
factor 2-3 (e.g. \cite{menci08}).

The exceptional high-resolution provided by the Advanced Camera for Surveys (ACS) at the 
Hubble Space Telescope (HST) has greatly contributed to the current knowledge on the evolution of 
galaxies in dense environments. Results on the eight z$\sim$1 clusters of the ACS Intermediate 
Redshift Cluster Survey (\cite{blake03}; \cite{mei07}, \cite{holden}, \cite{mei08} and references therein), 
and studies of individual distant clusters (RDCS 1252.9-2927 at $z$=1.235: \cite{lidman04},
\cite{demarco}; XMMU J2235.3-2557 at $z$=1.393: \cite{rosati09}, \cite{lidman08}; 
XMMXCS J2215.9-1738 at z=1.45, \cite{stan145}) point 
toward the prevalence of a tight RS up to $z$=1.4, where the CMR slope and scatter are observed 
to have a negligible increase with redshift. 

\begin{figure*}
\begin{center}
\includegraphics[width=10cm,angle=0,clip=true]{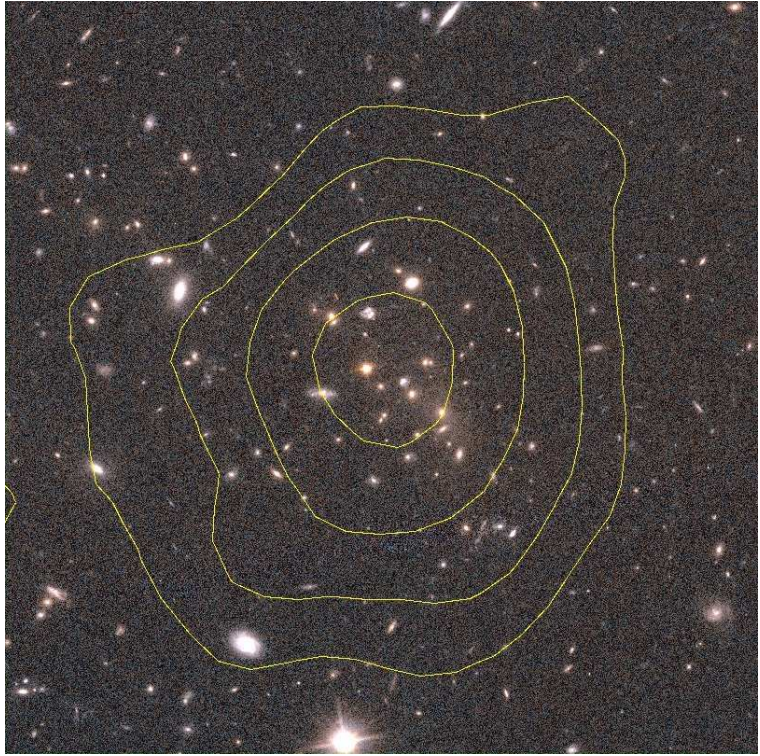}
\caption{HST/ACS colour image of XMMU J1229+0151 with X-ray contours. The image is centered on the 
cluster X-ray emission and has a size of 1.5 arcmin$^{2}$.}
\label{Figcolourima}
\end{center}
\end{figure*}

In this paper we provide a detailed analysis of the galaxy properties in XMMU J1229+0151 (hereafter, 
XMM1229), an X-ray selected, optically rich and distant cluster ($z$=0.975 corresponding to a lookback 
time of 7.6 Gyr). 
We derive accurate colour measurements from the high-resolution ACS data, and characterize 
the galaxy morphology via visual inspection and by fitting Sersic profiles. 
Stellar masses, ages and star formation histories of the cluster's early-types are derived 
by fitting the coadded spectrophotometric data with \cite{bruzual} templates.

The paper is organized as follows: in Sect. 2 we present the imaging and 
spectroscopic data, as well as reduction procedures. The ACS morphological analysis is 
introduced in Sect. 3. In Sect. 4 we derive the $i_{775}-z_{850}$ CMR, and the results 
from the SED fitting are presented in Section 5. In Sect. 6 we investigate the properties of 
the brightest cluster galaxies. We conclude in Sect. 7.

The cosmological parameters used throughout the paper are $H_{0}$=70 km/s/Mpc, $\Omega_{\Lambda}$=0.7 
and $\Omega_{\rm m}$ =0.3. Filter magnitudes are presented in the AB system unless stated otherwise.


\section{Observations and data reduction}

\subsection{XMM-Newton data}

The cluster XMM1229 was initially detected in a serendipitous cluster survey of the XMM-Newton 
archive, the XMM-Newton Distant Cluster Project (XDCP, \cite{boehringer05}, \cite{fasst}). 
Our target was observed in 25 XMM-Newton pointings of the bright radio loud quasar 3C~273 at an 
off-axis angle of approximately 13~arcmin. We selected only observations
whose exposure time, after cleaning for high background periods, was larger than 10~ks.
Unfortunately, XMM1229 was not observed by the EPIC-pn camera, since the pn was always 
operated in Small Window Mode (except for Obs\_Id=0126700201, having a clean exposure time of 
only $\sim6$~ks). Therefore, we used only the data from the two XMM/MOS CCDs.
The 11 observations selected for our analysis are listed in Table~\ref{log-xmm}. The information 
given is the following: observation date (column 1), XMM-Newton observation 
identification number (column 2) and revolution (column 3), filter (M=medium, T=thin) 
and mode (F=full window, S=small window) used (column 4), good exposure time 
of XMM/MOS1+MOS2, after cleaning for high particle background periods (column 5).

Data were processed using the XMM-Newton Science Analysis Software ({\tt SAS v7.0.0}).
Light curves for pattern=0 events in the $10-15$~keV band were produced 
to search for periods of background flaring, which were selected and removed by applying a 
3$\sigma$ clipping algorithm. 
Light curves in the $0.3-10$~keV band were visually inspected to remove residual 
soft-proton induced flares. 
We selected events with patterns 0 to 12 (single, double and quadruple) and further 
removed events with low spectral quality (i.e. {\tt FLAG=0}).
We obtained total exposure times of $\sim370$ and $\sim400$~ks for the 
XMM/MOS1 and XMM/MOS2, respectively.

The spectra of the cluster were extracted from a circular region of radius $30$~arcsec
centered at RA=12:29:29.2, Dec=+01:51:26.4.
The background was estimated from a circular region on the same chip of radius $\sim2$~arcmin 
centered at RA=12:29:21.2, Dec=+01:51:55.4, after removing cluster and point sources.

We corrected vignetting effects using the {\tt SAS} task {\tt EVIGWEIGHT} 
(\cite{arnaud01}) to calculate the emission weighted effective area, by 
assigning a weight to each photon equal to the ratio of the effective area at 
the position of the photon with respect to the on-axis effective area.
Redistribution matrices were generated using the {\tt SAS} task {\tt RMFGEN} 
for each pointing, filter, and detector.

Time averaged spectra for the source and the background were obtained by adding 
the counts for each channel. Since different filters were used for the observations, 
we weighted each instrument effective area (ARF) and redistribution matrix (RMF) with 
the exposure time of the observation.
In order to use the $\chi^2$ minimization in the spectral fitting we binned the spectra 
with a minimum number of 20 counts per bin.

\begin{table}
\caption{Log of the archival XMM-Newton observations of XMM1229.}
  \begin{center}
     \begin{tabular}{lllll}

    \hline\hline 

\bf{Date}    &     \bf{Obs. Id.} & \bf{Rev.} & \bf{Filt./Mode} & \bf{T$_{exp}$ [ks]}  \\
(1)          &     (2)           & (3)       & (4)       & (5)      \\
     \hline
2000-06-13   & 0126700201 &  0094  &  M/F       &  11.7+11.6        \\
2000-06-14   & 0126700301 &  0094  &  M/F       &  56.4+56.1        \\
2000-06-15   & 0126700601 &  0095  &  M/S       &  24.0+23.7        \\
2000-06-16   & 0126700701 &  0095  &  M/S       &  17.5+17.8        \\
2000-06-18   & 0126700801 &  0096  &  M/S       &  40.8+41.1        \\
2001-06-13   & 0136550101 &  0277  &  T/S       &  40.1+40.1        \\
2003-07-07   & 0159960101 &  0655  &  T/S       &  51.3+54.6        \\
2004-06-30   & 0136550801 &  0835  &  T1-M2/S   &  14.3+47.7        \\
2005-07-10   & 0136551001 &  1023  &  M/S       &  26.9+26.7        \\
2007-01-12   & 0414190101 &  1299  &  M/S       &  57.3+55.5        \\
2007-06-25   & 0414190301 &  1381  &  M/S       &  26.8+26.2        \\
       \hline

    \label{log-xmm}
    \end{tabular}
  \end{center}
\end{table}

\subsubsection{Spectral analysis}

The two XMM/MOS spectra were analyzed with XSPEC v11.3.1 (\cite{arnaud}) and 
were fitted with a single-temperature {\tt mekal} model (\cite{kaastra}; 
\cite{liedahl}). We modeled Galactic absorption with {\tt tbabs} (\cite{wilms}).
We always refer to values of solar abundances as in \cite{anders}.

The fits were performed over the $0.5-6$~keV band. We excluded energies below 
0.5~keV, due to calibration uncertainties, and above 6~keV, where
the background starts to dominate.
Furthermore, due to the relatively low S/N of the observations, 
we notice that instrumental K$\alpha$ emission 
lines\footnote{http://xmm.vilspa.esa.es/external/xmm\_user\_support/documentation
/uhb/node35.html} 
from Al (at $\sim1.5$~keV) and Si (at $\sim1.7$~keV) 
may affect the spectral analysis significantly. Therefore, we also excluded 
photons in the energy range $1.4-1.8$~keV from our spectral analysis.
In the selected energy bands we have a total of $\sim1300$ and $\sim1200$ 
net counts for the XMM/MOS1 and XMM/MOS2, respectively. 

The free parameters in our spectral fits are temperature, metallicity,
redshift and normalization, although we also performed the fit freezing the redshift to 0.975, 
the median spectroscopic redshift of the confirmed galaxies. Local absorption is fixed to the 
Galactic neutral hydrogen column density, as obtained from radio data (\cite{dickey}). 

\begin{figure}
\begin{center}
 \includegraphics[width=6.0cm,angle=-90]{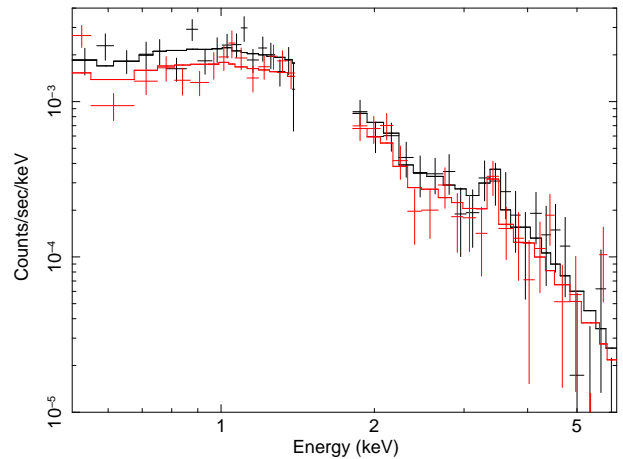}
 \caption{X-ray spectra of XMM1229 from the XMM/MOS1 (black) and XMM/MOS2 (red) detectors.
The solid lines show the best-fit models. The Fe K-line is prominent at 3.2 keV.}
 \label{xray-spec}
\end{center}
\end{figure}

\begin{table}[h]
\caption{Results of the X-ray spectral analysis. }
  \begin{center}
     \begin{tabular}{lllll}

    \hline\hline 

\bf{Detector} & \bf{kT [keV]} & \bf{Z$_{\mathrm{Fe}}$ [Z$_{\odot}$]} & \bf{z} & \bf{$\chi^2$/d.o.f.}  \\
(1)           & (2)           & (3)                                  & (4)    & (5)      \\
     \hline
MOS1    & $6.2_{-0.8}^{+1.0}$ & $0.37_{-0.18}^{+0.20}$ & $0.96_{-0.03}^{+0.02}$ & $45.1/51$      \\
MOS2    & $6.2_{-0.6}^{+1.1}$ & $0.42_{-0.21}^{+0.22}$ & $0.95_{-0.02}^{+0.03}$ & $50.1/44$      \\
MOS1+2  & $6.25_{-0.55}^{+0.69}$ & $0.38\pm0.14$       & $0.96\pm0.02$           & $95.4/98$      \\
MOS1+2  & $6.4_{-0.6}^{+0.7}$ & $0.34_{-0.13}^{+0.14}$ & $0.975^{\mathrm1}$     & $96.8/99$      \\      
		\hline
    \label{xmm-results}
\end{tabular}
\begin{list}{}{}
\vspace*{-0.5cm}
\item$^{\mathrm1}$ fixed redshift
\end{list}
  \end{center}
\end{table}

The results of the spectral analysis are listed in Table~\ref{xmm-results}, where the quoted 
errors are at the 1-$\sigma$ confidence level.
The information provided is the following: detector used, (column 1), temperature 
(column 2), iron abundance (column 3), redshift (column 4), $\chi^2$ and number of 
degrees of freedom (column 5). The last line of the table refers to the spectral fit 
with the redshift set to 0.975.

The rest-frame luminosity corrected for Galactic absorption in the $0.5-2.0$ keV range is  
$(1.3\pm0.2)\times10^{44}$~erg~s$^{-1}$, for an aperture of 30 arcsec radius, which 
corresponds to a physical size of $~$240 kpc. 
To obtain the total cluster luminosity we resort to extrapolating the measured aperture 
luminosity to a radius of 2 Mpc, assuming an isothermal $\beta$-model (\cite{cavaliere}), 
a well-known analytical formula dependent on a slope $\beta$ and a core radius $r_{c}$, that 
describes to a good degree the surface brightness profile of regular clusters.
Unfortunately, the signal-to-noise ratio is not good enough to fit a $\beta$-model, 
therefore we assume that XMM1229 is a cluster with a standard X-ray morphology i.e., 
without signs of merging or a strong cool core, and we use the typical values $\beta$=0.7 
and $r_{c}$=250 kpc, obtaining L$_{X}$ (r$<$2 Mpc) $\sim 3\times10^{44}$~erg~s$^{-1}$.

\subsection{HST/ACS $i_{775}$ and $z_{850}$ band imaging}

In the framework of the Supernova Cosmology Project (\cite{dawson}) we obtained 
the HST/ACS Wide Field Camera (WFC) data. Images in the F775W ($i_{775}$) and F850LP ($z_{850}$) 
passbands centered on the cluster X-ray centroid were acquired in December 2006, with total 
exposures of 4110 sec and 10940 sec respectively.
The $i_{775}$ and $z_{850}$ are the most efficient filters in supernova searches and, although they 
are not optimal for a cluster at this redshift, the $i_{775}$ encloses the D4000 break, which is 
redshifted to 7920 \AA~at the cluster redshift.
ACS has a field of view (FoV) of 3.3 x 3.3 arcmin and a pixel scale of 0.05"/pix.
The images were processed using the ACS GTO $Apsis$ pipeline (\cite{blake03b}), with a $Lanczos3$
interpolation kernel.
The photometric zero points are equal to 34.65 and 34.93 in the $i_{775}$ and $z_{850}$ bands 
respectively, following the prescription of \cite{sirianni}.
To account for the galactic extinction we applied to our photometric catalog the correction 
factor $E(B-V)$=0.017 
retrieved from the NASA Extragalactic Database\footnote{http://nedwww.ipac.caltech.edu/} 
and using the dust extinction maps from \cite{schlegel}. The corresponding correction in 
the optical bands is E($i_{775}$-$z_{850}$)=0.010 mag.

\subsection{VLT/FORS2 spectroscopy}

Spectroscopic observations were carried out with Focal Reducer and Low
Dispersion Spectrograph (FORS2: \cite{appen}) on Antu (Unit
1 of the ESO Very Large Telescope (VLT)) as part of a program to search
for very high redshift Type Ia supernova in the hosts of early type
galaxies of rich galaxy clusters (\cite{dawson}). In this
respect, the field XMMU~J1229+0151 was very rich in candidates, with
three candidates occurring during the three months monitoring. One candidate
was identified as a Type Ia supernova at the cluster redshift (\cite{dawson}).

FORS2 was used with the 300I grism and the OG590 order sorting
filter. This configuration has a dispersion of 2.3 Angstroms per pixel
and provides a wavelength range starting at 5900 \AA~and
extending to approximately 10000 \AA.  Since the observations
had to be carried out at short notice (the SN had to be observed while it
was near maximum light), most of the observations were done with the
multi-object spectroscopic (MOS) mode of FORS2. The MOS mode consists
of 19 moveable slits (with lengths that vary between 20" and 22") that
can be inserted into the focal plane. The slit width was set to 1". On
one occasion, when the MOS mode was unavailable because of technical
reasons, the field was observed with the MXU mode of FORS2. The MXU
mode consists of precut mask that is inserted into the FORS2 focal
plane once the field has been acquired. Since the length of the slit
can be made much shorter in the MXU mode than in the MOS mode, the
number of targets that could be observed in the MXU mode was a factor of
two larger than the number of targets that could be observed in the
MOS mode. However, the time to prepare, cut and insert a mask is
usually a couple of days, whereas the MOS observations can be done
with a few hours notice.

The field of XMMU J1229+0151 was observed with four different
configurations, 4 MOS and one MXU. The details of the
observations are given in Table~\ref{forslog}. The MOS configurations were 
used when the supernovae (there were three supernova visible in the field
of XMMU J1229+0151 at the same time) were near maximum light. The MXU
mask was used several months later when the supernovae were
significantly fainter. In all masks, slits were placed on the
supernova, thus spectra of the supernovae together with their hosts
and spectra of the hosts without the supernovae were obtained. The
other slits were placed on candidate cluster galaxies or field
galaxies. For each MOS setup, between 4 and 9 exposures of 700 to 900
seconds were taken. Between each exposure, the telescope was moved a
few arc seconds along the slit direction. These offsets, which shift
the spectra along detector columns, allow one to remove detector
fringes, as described in \cite{hilton}, which also details how
the FORS2 data was processed.

\begin{figure}
\begin{center}
 \includegraphics[width=7.5cm,height=5.5cm,angle=0,clip=true]{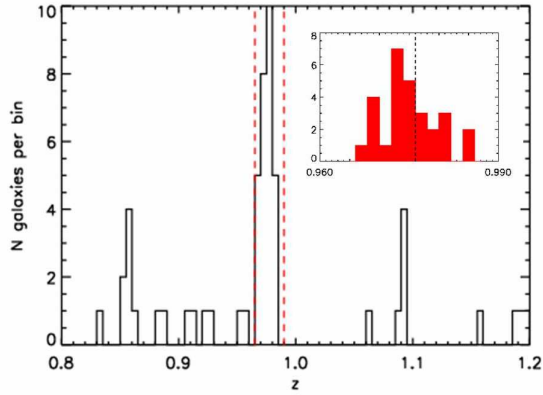}
 \caption{Redshift distribution of the galaxies in the cluster XMMU J1229+0151.
Vertical red-dashed lines refer to redshift cuts at $z$=0.965, 0.990 used to 
select the cluster members. This region is shown in more detail in the top-right inslet.}
 \label{Figredist}
\end{center}
\end{figure}

A total of 100 slits over four masks were used to observe 74 individual targets. 
The targets were selected by colour and magnitude, using the R- and z-band 
pre-imaging. Priority 1 targets had (R-z)$>$1.8 and z$<$23. Priority 2 targets had 
1.8$<$(R-z)$<$1.6 and z$<$23. 
Some cluster members were observed in more than one mask. From these
74 targets, 64 redshifts were obtained, and 27 of these are cluster members - 
 the redshift distribution of the targets in shown in Fig.~\ref{Figredist}.
A total of 21 confirmed galaxies are within the FoV of ACS.

\begin{table}
\caption{FORS2 Observing Log}
  \begin{center}
{\tiny
     \begin{tabular}{lllllll}
    \hline\hline 
\bf {Mask} & \bf{Type} & \bf{Slits} & \bf{Grism \& Filter}  & \bf{T$_{exp}$ }  & \bf{Airmass}  & \bf{Date}    \\
     &              &                &      &  (s)             &               &  (UT)        \\
     \hline
1   &  MOS  &	12 &	300I+OG590	&	8 x 750	&	1.3	&	2006 Jan 01      \\
2   &  MOS  &	18 &	300I+OG590	&	8 x 700	&	1.2	&	2006 Jan 30      \\
3   &  MOS  &	18 &	300I+OG590	&	4 x 700	&	1.1	&	2006 Jan 31      \\
4   &  MOS  &	18 &	300I+OG590	&	4 x 700	&	1.2	&	2006 Feb 01      \\
5   &  MXU  &	34 &	300I+OG590	&	9 x 900	&	1.3	&	2006 Jun 20-21   \\
      \hline
    \label{forslog}
    \end{tabular}
}
  \end{center}
\end{table}

Cluster membership was attributed by a reasonable selection of galaxies within 
$\pm$ 2000 km/s relative to the peak of the redshift distribution, or $\sim$ 3-$\sigma$. 
We assign the mean value of the redshift distribution of the 27 cluster members to the 
cluster redshift, $z$=0.975 and assume a conservative redshift error $\Delta z$=10$^{-3}$.
The cluster velocity dispersion was determined with the 27 galaxy redshifts, using the 
software ROSTAT of \cite{beers}. We obtained $\sigma$=683$\pm$62 km/s, where the error refers 
to the formal bootrap error obtained with 1000 samples. This value is in perfect 
agreement with the result obtained using the methodology proposed by \cite{danese}. 

Even though we have a limited number of cluster members which could introduce a bias 
in our computation of $\sigma$ due to the presence of substructures we, nevertheless, investigate 
the connection between the state of the hot intra-cluster medium (ICM) and the cluster galaxy population 
by means of the well-known Temperature-$\sigma$ relation (e.g. \cite{wu}). 
The observed T-$\sigma$ relation for high-$z$ clusters (e.g. \cite{rosati02}) implies that 
we would expect a higher velocity dispersion of about 900$\pm$300 km/s for the cluster temperature. 
We note however that there is a significant scatter in the T-$\sigma$ relation, and our value 
is within the 30\% scatter.

In Table~\ref{spec} we list the cluster members. The information provided is the following: 
galaxy ID (column 1); RA (column 2) and DEC (column 3); redshift (column 4); spectral 
classification (column 5) and morphological type (column 6).
The spectral classification is done according to the scheme proposed in \cite{dressler}, 
based on the strength of the [OII] and H$_{\delta}$ lines. The k class refers to passive 
(no [OII] emission) galaxies. This class is subdivided in two types, depending on the strength 
of the H$_{\delta}$ lines: k+a have moderate (3 $<$ EW H$_{\delta}$ $<$ 8) H$_{\delta}$ absorption, 
and a+k show strong (EW H$_{\delta}$ $>$ 8) H$_{\delta}$ absorption. 
The e spectral class refers to galaxies with [OII] emission and is subdivided into three types: 
e(a) present strong Balmer absorption, e(c) have weak or moderate Balmer absorption, 
and e(b) show very strong [OII] lines.

\begin{table}
\caption{Spectroscopic confirmed members.}
  \begin{center}
     \begin{tabular}{llllll}

    \hline\hline 

\bf{ID} & \bf{RA (J2000)} & \bf{DEC (J2000)}  & \bf{z}  & \bf{Class} & \bf{Type} \\
(1)     & (2)             & (3)               & (4)    & (5)       & (6) \\
     \hline
5417	& 187.3857875	& 1.8712528	& 0.977 &  a+k  & S0	\\
3428	& 187.3793000	& 1.8563222	& 0.984 &  a+k  & S0	\\
3430	& 187.3720375	& 1.8560639	& 0.974 &  k    & Ell	\\
3025	& 187.3771333	& 1.8363889	& 0.979 &  e(c) & Ell	\\
4055	& 187.3573750	& 1.8601056 	& 0.968 &  k    & Sb	\\
3301	& 187.3466250	& 1.8502667	& 0.969 &  e(c) & Ell	\\
4155	& 187.3885500	& 1.8644889	& 0.969 &  k+a  & Ell	\\
5411	& 187.3718958	& 1.8717778	& 0.974 &  k    & Ell	\\
20008	& 187.3734583	& 1.8726667	& 0.973 &  e(a) & Irr	\\
3497	& 187.3724875	& 1.8579083	& 0.982 &  a+k  & Ell	\\
20010	& 187.3726625	& 1.8579944	& 0.977 &  k    & Ell	\\
4126	& 187.3900292	& 1.8628750	& 0.973 &  k    & Ell	\\
3507	& 187.3716250	& 1.8571444	& 0.976 &  k    & Ell	\\
20013	& 187.3684167	& 1.8559167	& 0.979 &  k    & Ell	\\
20014	& 187.3654583	& 1.8485556	& 0.969 &  k    & S0	\\
3949	& 187.3696083	& 1.8602611	& 0.976 &  k    & Ell	\\
30004	& 187.3697875	& 1.8601389	& 0.970 &  k    & S0/Ell\\
3495	& 187.3715708	& 1.8582111	& 0.980 &  k    & Ell	\\
3524	& 187.3807000	& 1.8676667	& 0.969 &  a+k  & Ell	\\
5499	& 187.3844542	& 1.8683722	& 0.973 &  k    & Ell	\\
3205	& 187.3747292	& 1.8461806	& 0.984 &  a+k  & Ell	\\
4661$^{\mathrm a}$	& 187.3631292	& 1.8977194	& 0.975 &  k+a  & --	\\
5001$^{\mathrm a}$	& 187.3500083	& 1.8870944	& 0.973 &  k    & --	\\
4956$^{\mathrm a}$	& 187.3367208	& 1.8890611	& 0.978 &  k    & --	\\
4794$^{\mathrm a}$	& 187.3342500	& 1.8927333	& 0.974 &  k    & --	\\
4910$^{\mathrm a}$	& 187.3213042	& 1.8925528	& 0.976 &  e(b) & --	\\
4800$^{\mathrm a}$	& 187.3186708	& 1.8948944	& 0.976 &  k    & --	\\
      \hline
    \label{spec}
    \end{tabular}
\begin{list}{}{}
\vspace*{-0.5cm}
\item$^{\mathrm a}$ galaxy outside the FoV of ACS 
\end{list}
  \end{center}
\end{table}

\begin{figure*}
\begin{center}
 \includegraphics[height=11cm,angle=0,clip=true]{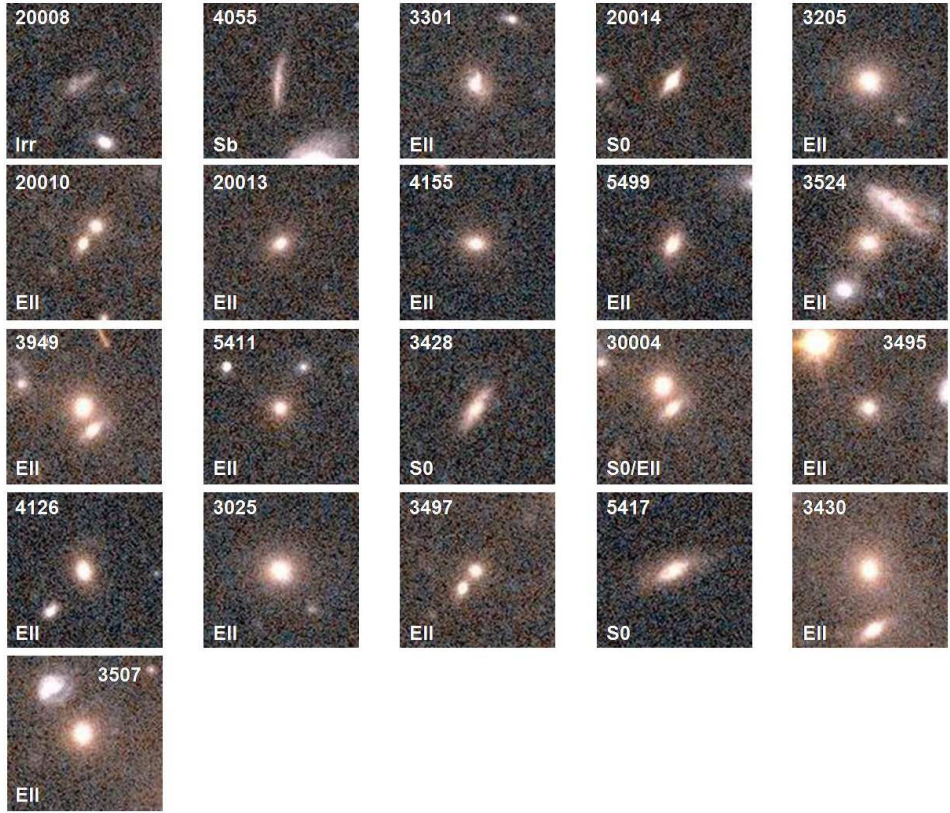}
 \caption{$i_{775}$-$z_{850}$ image gallery of the 21 spectroscopically confirmed members in the ACS 
FoV, ordered in increasing $i_{775}$-$z_{850}$ colour. Individual stamps are centered on the 
cluster members and have a size of 5" x 5". Top labels correspond to the spectroscopic galaxy 
ID and bottom labels refer to the visual morphological classification.}
 \label{Figstamps}
 \end{center}
\end{figure*}

\subsection{NTT/SOFI J- and Ks-band imaging }

NIR imaging in the J- and Ks-bands were acquired using SOFI (\cite{moor}) at 
the New Technology Telescope (NTT) at the ESO/La Silla observatory. The observations were taken 
in March 2007, as part of the NIR follow-up of the XDCP survey programme. The instrument was 
operated in the \textit{Large field} mode, corresponding 
to a 5x5 arcmin FoV, with a pixel scale of 0.288 arsec/pix. Since the NIR background 
is generally highly variable, a large dithering pattern has to be applied, thus we set the automatic jitter 
box to a width of 30 arcsec. Total exposure times amount to 1hr in Ks and 45 min in J. The J-band data 
have a seeing of 0.98" whereas the Ks-band have an image quality of 0.69".

Photometric calibration standards (\cite{persson}) were acquired several times during 
the observation run. The zero points (ZP) were computed using the reduced standards (background 
subtracted, count rate image) with the following relation:
\begin{equation}
 ZP= mag + 2.5 log(count rate) + atm_{corr}*airmass
\end{equation}

\noindent where $mag$ refers to the standard star magnitude, and $atm_{corr}$ refers to the 
wavelength dependent atmospheric correction. The stellar flux was measured within circular 
apertures with 6" radius; such a large radius ensures that we account for the bulk of the 
flux. The background was estimated with a 3$\sigma$ clipping algorithm.
The scatter of the zero points is 0.015 mag and 0.04 mag for the J and Ks 
filter, respectively. We converted VEGA magnitudes to the AB photometric system 
with the ESO web-tool available at $http://archive.eso.org/apps/mag2flux$.

The data was reduced with the package ESO/MVM (\cite{vandame}) using the HST/ACS catalog to 
match the astrometry. We used SExtractor (\cite{bertin}) in \textit{dual image mode} 
to perform the source detection in the Ks-band, and the photometry of both images. 

\section{Structural analysis}

\subsection{Surface brightness profile fitting}

The radial surface brightness profiles of galaxies can be described by the Sersic law 
(\cite{sersic}), 
\begin{equation}
 \Sigma(r) \propto exp{(r/r_{e})^{1/n} -1 } 
\end{equation}
\noindent where $\Sigma(r)$ is the surface brightness at radius $r$, the Sersic index, $n$, 
characterizes the degree of concentration of the profile; and the effective radius, $R_{e}$, 
corresponds to the projected radius enclosing half of the galaxy light.

Using the ACS $i_{775}$ and $z_{850}$ data we made a 2D bulge/disk galaxy decomposition 
with the software GIM2D 
(\cite{simard}). The galaxy model is the sum of a bulge component (Sersic profile) 
and an exponential disk, depending on a total of eleven parameters. Of these parameters, three 
describe the shape of the Sersic profile, including the index $n$, which we constrained to 
0$<n<$4. The upper bound is introduced because $n$=4 corresponds to the de Vaucouleurs 
profile, a purely empirical fit to the profiles of elliptical galaxies and bulges (\cite{devauc}).
Allowing larger values of $n$ usually does not improve the fit, however the covariance 
between $n$ and $R_e$ can lead to an overestimation of $R_e$ for large $n$ (\cite{blake06}). 
The median Sersic index $n$ of the spectroscopically confirmed galaxies is 3.9 and the median 
effective radius is 5.5 pixel (0.28"). 

\begin{figure}
\begin{center}
\includegraphics[width=8.5cm,angle=0,clip=true]{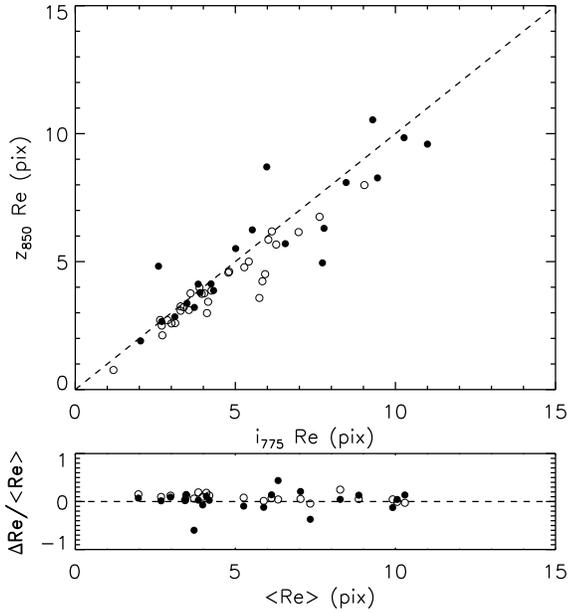}
 \caption{Comparison of the $R_{e}$ obtained with GIM2D in the i$_{775}$ and z$_{850}$ bands. 
 Spectroscopic members are represented in solid circles and "red-sequence" galaxies 
(see Sect. 4.2)) are shown in open circles. The dashed line indicates the one-to-one relation. 
The bottom plot shows the difference in the $R_{e}$ values for the two bands normalized by the 
average $R_{e}$. The dashed line represents the constant zero value.}
 \label{Figre}
\end{center}
\end{figure}

The distribution of $R_{e}$ is consistent in both bands within the 1-$\sigma$ errors, 
with an average error of 0.77 and 0.53 pix in the $i_{775}$ and $z_{850}$ band 
respectively. The comparison between the effective radii obtained in the two bands 
$R_{e}$ ($i_{775}$) - $R_{e}$ ($z_{850}$) is shown in Fig.~\ref{Figre}. 
This difference is useful to assess an imperfect matching of the PSFs or the 
presence of colour gradients. However, we find a very good agreement between the two 
radii therefore we do not expect those effects.
In this figure we also present the results of fitting a "red-sequence" sample 
of early-type galaxies which is introduced in Sect. 4.2.
The reduced $\chi^{2}$ of the best-fit models is $\sim$1 for the 
majority of the galaxies, emphasizing the good quality of the fit.

\subsection{Visual morphological classification}

\begin{figure}
\begin{center}
\includegraphics[width=7.5cm,angle=0,clip=true]{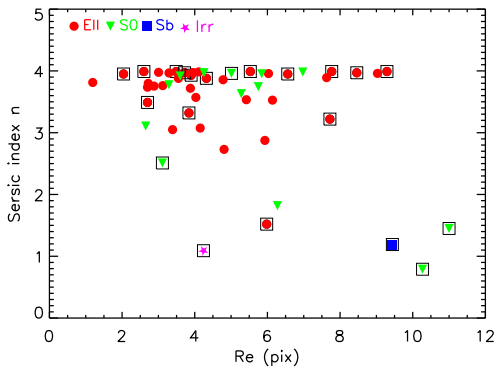}
 \caption{Sersic index $n$ as a function of the effective radius 
$R_{e}$ obtained with GIM2D, using the ACS/$i_{775}$ band. The spectroscopic sample is 
evidenced by open squares. The morphology of the cluster members is dominated by elliptical 
galaxies (red circles) characterized by a high $n$. Four galaxies are classified as S0 
(green triangles), one member is an Sb galaxy (blue square), with $n<1$ and one galaxy 
has an irregular shape (magenta 5-pointed star). The 31 "red-sequence" early-type 
galaxies (see Sect. 4.2) are also displayed with the same symbols without the open squares.}
 \label{Figmorph}
\end{center}
\end{figure}

In addition to the profile fitting, we made a visual classification of the spectroscopic 
members using morphological templates from \cite{postman}. In Fig.~\ref{Figstamps} we show postage 
stamps of the cluster members in the $i_{775}$ passband labelled with the morphological type. 
We note two red galaxy pairs (ID=20010/3497, 30004/3949). 
In Fig.~\ref{Figmorph} we show the distribution of the fit parameters $n$ and $R_e$ 
of both the spectroscopic and "red-sequence" samples (see Sect. 4.2 for details on the latter), 
complemented with the visual classification.

The morphology of the spectroscopic galaxies in XMM1229 is clearly dominated by elliptical galaxies 
(15/21) with only one galaxy classified as spiral (ID=4055) and one irregular (ID=20008), 
unlike other distant clusters (see for eg. the EDisCS high-redshift sample, $z \le$0.8, 
\cite{delucia04}). The remaining four cluster members are classified as S0s. 
We stress that we targeted red galaxies for spectroscopy, hence this was a colour, not 
a morphological selection, therefore we do not expect to have a bias on ellipticals with 
respect to S0s.

\section{The $i_{775}-z_{850}$ colour-magnitude relation}

\subsection{Galaxy photometry}

\begin{figure}
\begin{center}
 \includegraphics[width=6cm,angle=0,clip=true]{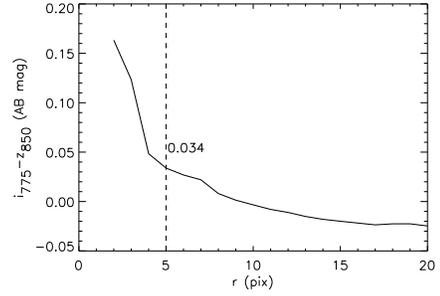}
 \caption{Differential PSF blurring effect in i and z-bands: at r=5 pix (0.25``)
 the PSF correction is 0.034 mag (vertical line). }
 \label{Figpsfcor}
\end{center}
\end{figure}

We use SExtractor in \textit{dual image mode} to perform the source detection in the 
$z_{850}$ band, and the photometry in both bands. 
The image quality of the $i_{775}$ band is sightly better than the $z_{850}$ band, 
with a Point Spread Function (PSF) FWHM of 0.085``, as opposed to 0.095'' in the $z_{850}$.
The effect of the $z_{850}$ PSF broadening has been investigated in other works 
(e.g. \cite{mei06}) and is attributed to the long-wavelength halo of the ACS/WFC (\cite{sirianni}).
This effect, although small, bears implications on the galaxy colour measurement and has to 
be accounted for. 
Thus, for each passband we constructed empirical PSFs by computing the median profile of a handful 
of stars in the science images for which we measure growth curves normalized to the central 
intensity. We obtain a differential ($z_{850}$-$i_{775}$) median radial profile that shows a 
steep behavior for radii smaller than 3 pix - see Fig.~\ref{Figpsfcor}.

The $i_{775}-z_{850}$ colour is determined in small apertures to avoid intrinsic galaxy colour 
gradients (see e.g. \cite{scode} for a discussion on the effect of internal colour gradients). 
We choose a fixed aperture of 5 pix (0.25'') since at 
this radius the steep and uncertain PSF broadening is no longer dominant 
(see Fig.~\ref{Figpsfcor}), and we apply a correction of 0.034 mag to the 
$i_{775}$ band in order to match the poorer seeing of the $z_{850}$ band. Total $z_{850}$ band 
magnitudes are obtained with SExtractor parameter {\tt MAGAUTO}.

\subsection{Colour-magnitude relation}

\begin{figure}
\begin{center}
 \includegraphics[width=9.5cm,angle=0,clip=true]{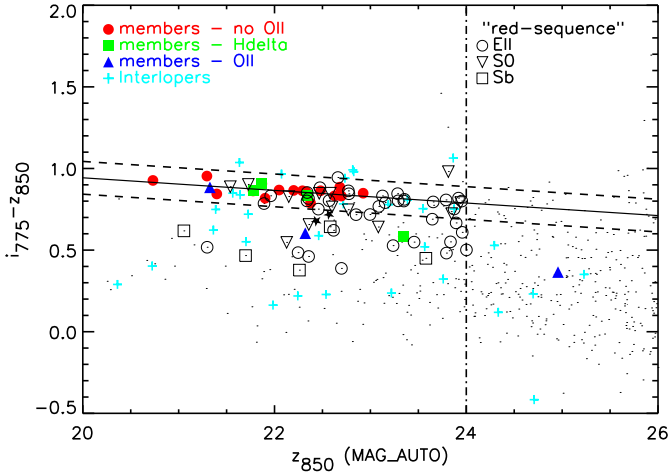}
 \caption{$i_{775}$-$z_{850}$ colour-magnitude relation of XMM1229. The black solid line 
refers to the linear fit to the passive cluster members, which are shown in red circles.
The dashed lines correspond to the 3-$\sigma$ region. 
The confirmed galaxies with [OII] emission are displayed in blue triangles; members with 
strong H$\delta$ absorption (see Sect. 5.2) are displayed in green squares. The cyan 
crosses refer to spectroscopically confirmed interlopers. The "red-sequence" sample 
is presented with open black symbols: circles refer to ellipticals, squares refer to 
S0s and inverted triangles correspond to Sb galaxies. A merging pair of elliptical galaxies 
is marked with two filled 5-pointed-stars. The black dots correspond objects within 1 arcmin 
from the X-ray cluster center.}
 \label{Figcmddata}
\end{center}
\end{figure}

The colour-magnitude relation is presented in Fig~\ref{Figcmddata}. We flag the 35 confirmed 
interlopers in the ACS field (cyan crosses), since nearly a fourth of them (9/35) are 
located on the red-sequence. 

We perform a robust linear fit using bi-square weights (Tuckey's Biweight) to the CMR of 
the confirmed passive members. 
The bi-square weights are calculated using the limit of 6 outlier-resistant standard 
deviations. The process is performed iteratively until the standard deviation changes by less 
than the uncertainty of the standard deviation of a normal distribution.
The linear fit has a slope of -0.039$\pm$0.013 and a zero point CMR$_{ZP}$=0.86$\pm$0.04, which 
was determined with a bi-weighted mean. The quoted uncertainty on the slope corresponds to 
the estimated standard deviation of the fit coefficient. The scatter of the CMR including 
only the passive galaxies is 0.039 mag. 

Since the spectroscopic sample does not populate well the faint end of the red-sequence, 
we selected a "red-sequence" sample, based on a combination of morphological and colour 
criteria. We applied a generous colour cut of 0.5 $< i_{775}-z_{850} <$ 1.3 for 20 $< z_{850} <$ 24, 
based on the properties of the bluest star forming cluster galaxies and the magnitude limit
 set by \cite{postman} to ensure a reliable morphological classification. In addition, we 
constrained the search radius to 1' from the cluster X-ray center, corresponding to 478 kpc at 
the cluster redshift. This is a reasonable area to search for cluster members, and avoids 
contamination of non-members. We can also express this radius as a fraction of the fiducial 
radius $R_{200}$ which was estimated using the $R_{200} - T$ X-ray scaling relations of 
\cite{arnaud05}. Thus, we determine the search radius of 478 kpc to be equal to 0.4$\times R_{200}$.

We found 58 galaxies in this region which were visually classified using the templates 
from \cite{postman}. 
The selection of the red-sequence galaxies was based on the 3-$\sigma$ clipping 
of the linear fit to the confirmed passive members. Thirty-one galaxies are the region 
delimited by the 3-$\sigma$ clipping, for a $z_{850}$ magnitude cut at 24 mag. Again the 
fraction of ellipticals, 22/31, is much larger than the fraction of S0s, 9/31.
The scatter of the red-sequence combining the two samples (spectroscopic and "red-sequence") 
is equal to 0.048 mag.

\subsection{Model colour-magnitude relation}

\begin{figure}
\begin{center}
 \includegraphics[width=8.5cm,angle=0,clip=true]{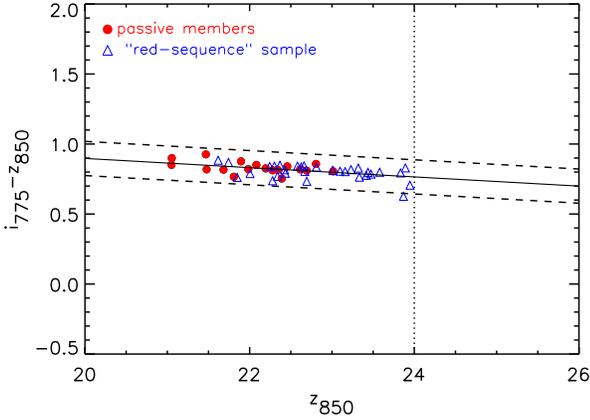}
 \caption{$i_{775}$-$z_{850}$ colour-magnitude relation of XMM1229 using the best-fit galaxy models. 
The black solid line refers to the composite linear fit to both the passive cluster members 
(red circles). The dashed lines delimit the 3-$\sigma$ region and the dotted line marks the 
$z_{850}$ magnitude cut at 24 mag.}
 \label{Figcmdimz}
\end{center}
\end{figure}

Traditionally, galaxy colours are measured either using aperture magnitudes with corrections 
which take into account PSF differences, or by using aperture magnitudes after deconvolving the PSF 
as, for e.g., in \cite{blake03}. 
Instead, in this work we explored a method to derive galaxy colours based on model magnitudes, as 
commonly used in the \textit{Sloan Digital Sky Survey} (e.g. \cite{blanton}).
In this method, the PSFs of the 2 filters (which are estimated independently in the two bands as 
described in the previous section) are convolved with the galaxy profile models. Hence, this is 
a direct method were convolution and $not$ deconvolution is performed.
We therefore use the surface brightness best-fit models with additional gaussian noise to 
measure aperture and total magnitudes. Similarly to the "data" CMR (Sect. 4.2), 
the colour measurements are performed in fixed apertures with r=5 pix.
An alternative approach would be to measure aperture magnitudes over the individual 
galaxy effective radius, however this strategy proved unreliable since for many 
galaxies $R_{e}$ is smaller than 3 pix (0.15"), which is really too small to make a 
proper colour measurement. Total magnitudes were derived using apertures with radius of 
10$\times R_{e}$ instead of using SExtractor {\tt MAGAUTO}, which we found to be inaccurate 
in comparison with large aperture magnitudes. These discrepancies are visible in the 
total $z_{850}$ magnitude of the brightest galaxies in the two CMR's represented in 
Figs.~\ref{Figcmddata} and \ref{Figcmdimz}.

The procedure to fit the CMR and obtain its zero point is identical to the method described 
earlier in Sect. 4.2, only that now we use both the early-type cluster members and the 
"red-sequence" sample. If we consider only the confirmed passive members to perform the 
linear fit we obtain a zero point CMR$_{ZP}$=0.83$\pm$0.04, a slope of -0.031$\pm$0.016 and a 
scatter equal to 0.042$\pm$0.011. The error of the scatter is estimated with 100 Monte Carlo 
simulations of the galaxy models, varying the Sersic index and the effective radius within the 1-$\sigma$ 
confidence errors. 
The uncertainty associated with the scatter is estimated by fitting a gaussian to the 
distribution of the scatter measured in the models and assigning the standard deviation of the 
distribution to the error.
In order to perform a composite linear fit to both the spectroscopic and "red-sequence" 
samples we applied a magnitude cut at $z_{850}$=24 mag, a limit that ensures a reliable 
visual classification of the "red-sequence" sample (see for e.g. \cite{postman}).
We obtain a CMR zero point equal to 0.81$\pm$0.04, the total intrinsic scatter slightly 
increases to 0.050 mag and the slope, -0.031$\pm$0.008, remains nearly unchanged.

We would like to remark that the scatter of the colour-magnitude relations derived 
from SAM is a factor 2-3 larger than the observational scatter (see for e.g. \cite{menci08}).  
In semi-analytical modelling the scatter is obtained by computing the total galaxy magnitudes, 
which is precisely known in simulations. A possible reason for the discrepancy between the 
observational scatter and the one obtained with simulations is the existence of colour gradients 
which are taken into account in the total galaxy magnitudes used in SAM to measure the scatter, 
whereas in the observations we limit the colour measurement to a small central aperture, thus 
minimizing the effect of such gradients.
To investigate this effect we measured the $i_{775}-z_{850}$ colour of the passive members using 
the galaxy models, increasing the fixed colour aperture to r=10,15 pix, respectively 0.5", 0.75" - 
going beyond these radii would produce noisy measurements since we would run into the background. 
The corresponding scatter is then 0.068, 0.088, respectively. This result suggests the presence 
of colour gradients in the galaxy sample.

\section{Analysis of the spectral energy distributions}

The observed spectral energy distribution of a galaxy is a record of its stellar population history.
The SED fitting method relies on the comparison of the observed SED with synthetic 
SED's. The latter are then convolved with the transmission curves of the filters used in the 
observations and the output magnitudes are compared with the observed magnitudes.
Galaxy SED's were determined by measuring the flux within a fixed aperture of 3 arcsec in 
the four available passbands.

Given the large disparity in the resolution of the ground- and space-based data, 
a careful matching of the different PSFs must be done, when constructing the multi-wavelength catalog 
for sampling the galaxies' SEDs. The method we used to derive aperture corrections is the following: 
we smoothed the $i_{775}$, $z_{850}$ and Ks-band images with gaussian kernels to match the seeing of 
the J-band ($\sim$1") and made growth curves of stars in the original and degraded (smoothed) images. 
We then obtained a differential median radial profile for each band with which we derive corrections 
at a given radius. 
In the multi-colour catalog we use galaxy magnitudes corrected to match the fluxes to the worst seeing 
image (J-band), measured within 0.5" radius apertures for the ACS bands, and 1.16" for the NIR data.
We opted to work with magnitudes extrapolated to 3" radius, which safely enclose the bulk of the 
galaxy flux.

A total of 20 spectroscopic members are common to the FoV of SOFI and ACS, four of which 
constitute two red galaxy pairs that are not properly resolved in the NIR data. 
For this reason we had to exclude them from the spectral analysis.
Only four of the remaining 16 studied galaxies show [OII] emission lines (IDs: 3025, 3205, 3301, 4055) 
signaling ongoing star formation, and the first two also present [OIII] lines. The first three 
galaxies have been visually classified as ellipticals, although galaxy 3301 has a low Sersic index, 
$n$=1.5. Galaxy ID=4055 is an edge-on spiral which is reflected in the low Sersic index ($n$=1.2).

Additionally, we also fitted the SEDs of the ETGs in the "red-sequence" sample lying on the 
ACS CMR red-sequence. As mentioned earlier, we find 31 ETG in the locus of the red-sequence. 
The poorer quality of the NIR data can only resolve 18 of these galaxies.

\subsection{Spectrophotometric properties: masses, ages}

\begin{figure}
\begin{center}
\includegraphics[width=7.5cm,angle=0,clip=true]{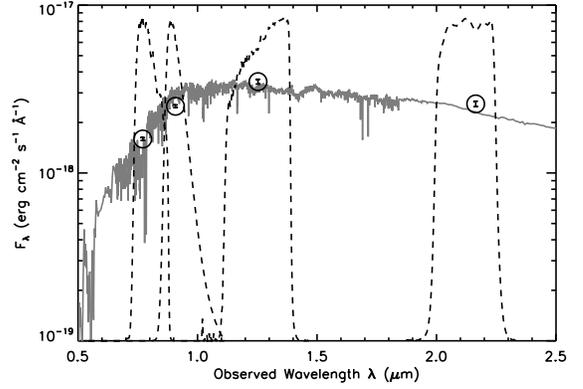}
\caption{SED fit of one of the three brightest galaxies, ID=3025 (solid line). The flux measurements 
in the $i_{775}$, $z_{850}$, J and Ks-bands (respectively from left-to-right) are shown in circles, 
with 1-$\sigma$ error bars, along with the filters transmission curves (dashed lines).}
\label{Figsed}
\end{center}
\end{figure}

Stellar masses, ages and star formation histories are derived from the synthetic galaxy fluxes, assuming 
solar metallicity and a Salpeter (\cite{salpeter}) initial mass function (IMF), with a mass cut off 
[0.1-100] M$_{\sun}$. We perform a three parameter (age T, $\tau$, mass) fit to the SEDs using a grid of 
\cite{bruzual} models characterized by a delayed exponential star formation rate: 
$\frac{t}{\tau^{2}}.exp(\frac{-t}{\tau})$, performing a minimization of the $\chi^{2}$. The 
parameter $\tau$ spans a range of [0.2- 5.8] Gyr, where 5.8 Gyr is the age of Universe at the 
cluster redshift. As an example, in Fig.~\ref{Figsed} we present the fit to the SED of 3025, 
one of the three brightest galaxies (see Sect.6), together with the filter transmission curves.

The star formation (SF) weighted age represents the mean age of the bulk of the stars in a 
galaxy (depending on the $\tau$ parameter), and is defined as: 
\begin{equation}
 t_{SFR}= \frac{\int_0^T dt'(T-t')\Psi(t')}{\int_0^T dt'\Psi(t')} 
\end{equation}
\noindent where $\Psi$ is the star formation rate 
expressed as 
\begin{equation}
\Psi= \tau^{-2} te^{\frac{-t}{\tau}} + A \delta (t-t_{burst}.)
\end{equation}
\noindent The parameter A refers to the amplitude of an instantaneous burst at 
time $t_{burst} > \tau$, as described in \cite{gobat}.
Galaxy SF weighted ages do not change significantly if other models (\cite{maraston}) 
and different IMF's are used (\cite{chabrier}, \cite{kroupa}), however the stellar masses 
are dependent on the chosen IMF.

\begin{figure}
\begin{center}
\includegraphics[width=8cm,angle=0,clip=true]{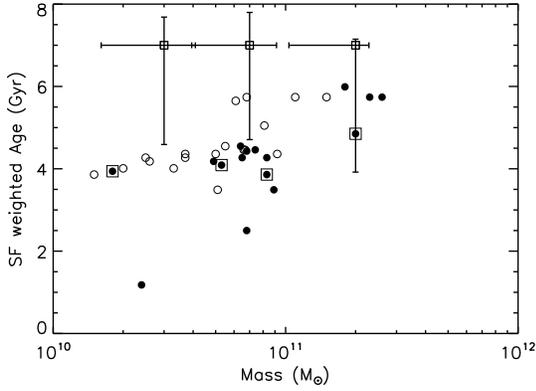}
\caption{Photometric masses of the 16 spectroscopic cluster galaxies (filled circles) and 18 
"red-sequence" ETGs (open circles) as a function of their star formation weighted ages. 
Spectroscopic members with [OII] emission or morphologically classified as late-type galaxies 
are signaled with square symbols. Mean error bars corresponding to three mass bins (bin1: m 
 $<$4.5$\times 10^{10}$ M$_{\sun}$, bin2: 4.5$\times 10^{10}$ M$_{\sun}$ $<$ m $<$1$\times 10^{11}$ 
M$_{\sun}$, bin3: m $>$1$\times 10^{11}$ M$_{\sun}$) are shown on the top.}
\label{Figmassage}
\end{center}
\end{figure}

\begin{figure}
\begin{center}
\includegraphics[width=8cm,angle=0,clip=true]{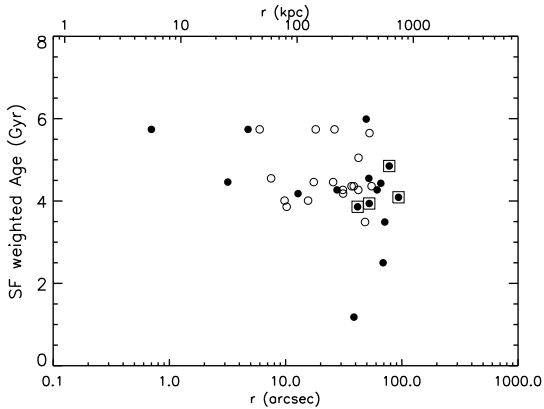}
\caption{SF weighted age \textit{versus} radial distance to the cluster center. Spectroscopic 
members with [OII] emission or morphologically classified as late-type galaxies are signaled with 
square symbols.}
\label{Figdaist}
\end{center}
\end{figure}

\begin{figure}
\begin{center}
\includegraphics[width=8cm,angle=0,clip=true]{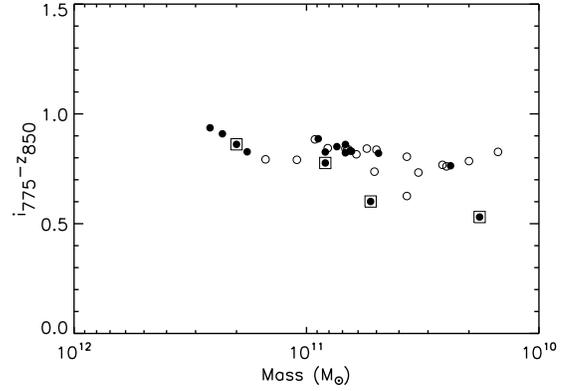}
\caption{Correlation between galaxy $i_{775}-z_{850}$ colour and photometric mass. Spectroscopic 
members with [OII] emission or morphologically classified as late-type galaxies are signaled with 
square symbols.}
\label{Figcolm}
\end{center}
\end{figure}

\begin{table}
\caption{SED analysis of the spectroscopic members.}
  \begin{center}
     \begin{tabular}{lllll}

    \hline\hline 

\bf {Galaxy ID} & \bf {Mass ($10^{10}$ M$_{\sun}$)} & \bf {t$_{SFR}$ Age (Gyr)} \\
(1)   		& (2)                    & (3)         \\
     \hline
3301	&	5.3$^{+1.8}_{-3.1}$	&	4.09$^{+0.34}_{-2.82}$	\\
4055	&	1.8$^{+1.1}_{-1.3}$	&	3.94$^{+0.58}_{-3.14}$	\\
20014	&	2.4$^{+2.2}_{-0.3}$	&	1.18$^{+1.34}_{-0.1}$	\\
20013	&	4.9$^{+0.8}_{-2.0}$	&	4.18$^{+0.18}_{-2.19}$	\\
5411	&	6.4$^{+2.3}_{-2.7}$	&	4.55$^{+1.19}_{-2.15}$	\\
3507	&	26$^{+2}_{-11}$		&	5.74$^{+0}_{-2.50}$	\\
3495	&	7.4$^{+3.6}_{-2.8}$	&	4.46$^{+1.28}_{-2.06}$	\\
3430	&	23$^{+3}_{-11}$		&	5.74$^{+0}_{-2.84}$	\\
3205	&	8.3$^{+0.3}_{-2.3}$	&	3.86$^{+0.15}_{-1.56}$	\\
3524	&	18$^{+2}_{-11}$		&	5.99$^{+0}_{-3.80}$	\\
3428	&	6.5$^{+1.0}_{-2.7}$	&	4.27$^{+0.19}_{-2.17}$	\\
3025	&	20$^{+6}_{-9}$		&	4.85$^{+0.89}_{-2.36}$	\\
5417	&	8.9$^{+5.1}_{-1.8}$	&	3.49$^{+2.25}_{-0.64}$	\\
5499	&	8.3$^{+3.7}_{-3.7}$	&	4.27$^{+0.68}_{-2.28}$	\\
4155	&	6.8$^{+0.9}_{-2.7}$	&	4.43$^{+0.18}_{-2.24}$	\\
4126	&	6.8$^{+1.3}_{-0.8}$	&	2.50$^{+0.91}_{-0.31}$	\\
      \hline
    \label{sed}
    \end{tabular}
  \end{center}
\end{table}

The spectroscopic cluster members form a fairly old population, with a median SF weighted age of 4.3 Gyr, 
and with stellar masses in the range 4$\times 10^{10}$-2.3 $\times 10^{11}$ M$_{\sun}$, 
see Table~\ref{sed} for the listing of the fitted masses (column 2) and ages (column 3).
The "red-sequence" sample, which allows us to probe fainter galaxies, appears to be less massive, 
with a median stellar mass of 5.5 $\times 10^{10}$ M$_{\sun}$. However, since we do not have 
redshifts for these galaxies, we cannot draw strong conclusions about their masses and formation ages.

In Fig.~\ref{Figmassage} we investigate the correlation between the star formation weighted age 
and stellar mass content in both the spectroscopic (filled circles) and "red-sequence" 
samples (open circles). We find a strong mass-age correlation which is confirmed with a Spearman 
$rho$ rank of 0.61 with a significance $p$ of 1.4$\times$10$^{-4}$ 
($p$ is a value lying in the range [0.0 - 1.0], where $p$=0 indicates a very 
significant correlation and $p$=1 means no correlation).
This mass-age correspondence evidences a well-known anti-hierarchical behavior ($downsizing$), 
where the most massive galaxies are also the oldest. 

We also investigate the dependence of the galaxy radial distance to the cluster center with 
mass and SF weighted age. We find that the most massive elliptical galaxies populate the cluster 
core, and conversely the four late-type galaxies are situated at the periphery of the cluster, 
at about 1 arcmin from the center, indicating star formation taking place in these regions. 
This morphological segregation is well established at lower redshifts (e.g. \cite{biviano}), 
nonetheless it is interesting to note that at redshift $\sim$ 1, the late type galaxies are 
already settled at the outskirts of the cluster. This segregation was also found by 
\cite{demarco05} and \cite{homeier} in the study of a cluster with z=0.837, as well as in 
RDCS 1252.9-2927 at z=1.234 (\cite{demarco}).
The dependence of the star formation weighted 
age with the cluster centric distance (Fig.~\ref{Figdaist}) shows that the galaxy age scatter 
increases at larger radii. This is indicative of younger/more diverse SF histories for galaxies 
located in the outer regions of the cluster, which have presumably accreted later onto the cluster.
This result has also been found in other work, e.g. \cite{mei08}.

Finally, we analyze the relation between the $i_{775}-z_{850}$ colour and the mass (Fig.~\ref{Figcolm}) 
and we observe the expected trend of the most massive galaxies being redder.

\subsection{Star formation histories}

\begin{figure}
\begin{center}
\includegraphics[width=8cm,angle=0,clip=true]{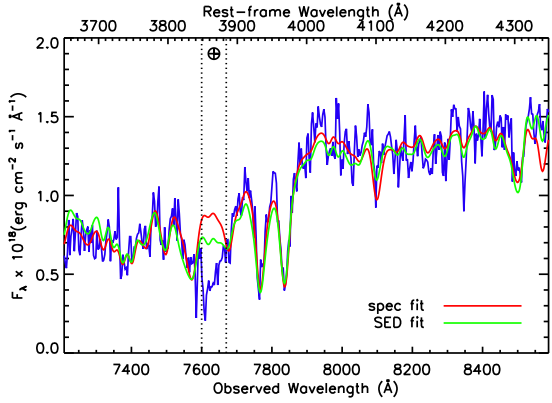}
\caption{Stacked spectrum (in blue) of all passive members which do not show strong H$_{\delta}$ 
absorption (k type, see Table~\ref{spec}). The red line refers to the best-fit model 
to the stacked spectrum and the green line refers to the best-fit model to the average SED. 
The region around the atmospheric A-band at 7600 \AA~(dashed lines) is difficult to subtract and 
was therefore ignored in the fit.}
\label{Figstack}
\end{center}
\end{figure}

The spectra of 12 confirmed passive members were coadded to obtain the stacked spectrum. 
However, four (ID=3428, 3524, 4155 and 5417) of these 12 galaxies have strong H$_{\delta}$ 
absorption, EW(H$_{\delta}$) $\ge$ 3 (these are a+k/k+a spectral types, see Table~\ref{spec}) 
and therefore we removed them from the stacking procedure. Three of these galaxies 
are at about 1 arcmin from the X-ray cluster center and only galaxy (ID=3428) is closer to the 
core, at $\sim$ 0.5 arcmin from the center. 
In Fig.~\ref{Figstack} we present the coadded spectrum of 8 spectroscopic passive members 
with weak or no H$_{\delta}$ absorption. 
The best-fit SED is shown in green and the spectral fit is shown in red. 

Star formation histories were derived only for the eight galaxies which do not have 
significant H$_{\delta}$ absorption.
The star formation weighted age and formation redshift obtained by the best fitting models 
(i.e. those within the 3-$\sigma$ confidence), is 3.7 $^{+0.4}_{-0.5}$ Gyr and 
$z_{f}$= 3.0 $\pm$ 0.5 respectively, when using the combined spectrophotometric data. 
It is not surprising that there is discrepancy between the average age obtained by fitting 
the individual SEDs (4.3 Gyr) and that derived from the composite spectrophotometric data, as we are 
using the spectrum and SED to place complementary constraints on the star formation histories (the 
former has resolution but poor wavelength coverage, while the latter has coverage but poor 
resolution). This discrepancy can stem from the fact that the SED unfortunately does not probe 
the rest-frame UV and would be thus somewhat insensitive to recent star formation.

\section{Is there a Brightest Cluster Galaxy?}

The cores of rich galaxy clusters most often host a massive and bright giant elliptical 
galaxy - the brightest cluster galaxy (BCG). 
In XMM1229, instead of one prominent BCG, we find three bright galaxies within $\sim$ 0.5 mag.
The total z$_{850}$-band magnitudes are derived by integrating the best-fit surface brightness 
model to a large radius, r=10$\times R_{e}$. In Table~\ref{tbcg} we summarize the most relevant 
characteristics of these galaxies: total z$_{850}$ magnitude from best-fit model (column 2), distance to the 
X-ray cluster center (column 3), photometric mass (column 4), star formation weighted age (column 5).
As expected, the three bright galaxies are the most 
massive galaxies, with masses of the order of 2 $\times 10^{11}$ M$_{\sun}$. The galaxy ID=3025 
located at 1.3`` from the cluster center shows strong [OII] emission, indicating ongoing star 
formation which is confirmed by a lower star formation weighted age of 4.85 Gyr, approximately 
1 Gyr younger than the other two brightest galaxies. In addition, this galaxy is fainter by $\sim$ 
0.2 mag in Ks and $\sim$ 0.15 mag in J, with respect to the other two bright galaxies.

\begin{table} [h]
\caption{Properties of the three brightest galaxies.}
  \begin{center}
     \begin{tabular}{lllll}
    \hline\hline 
\vspace{-0.3cm}
\\
\bf{ID} & \bf{z$_{850}$ mag} & \bf{dist ["]} & \bf{Mass [$10^{11}$ M$_{\sun}$]} & \bf{Age [Gyr]}  \\
(1)     & (2)       & (3)       & (4)       & (5)      \\
     \hline
3025    & $21.051 \pm 0.002$    & $78$ & $2.0_{-0.9}^{+0.6}$  & $4.9_{-2.36}^{+0.89} $      \\
3430    & $21.055 \pm 0.002$    & $5$  & $2.3_{-1.1}^{+0.3}$ & $5.7_{-2.84} $      \\
3507    & $21.468 \pm 0.002$    & $1$  & $2.6_{-1.1}^{+0.3}$ & $5.7_{-2.50} $      \\
     \hline
    \label{tbcg}
    \end{tabular}
  \end{center}
\end{table}


\section{Discussion and conclusions}

XMMU J1229+0151 is a rich, X-ray luminous galaxy cluster at redshift $z$=0.975, that 
benefited from a good multi-wavelength coverage and is therefore an adequate 
laboratory for studying galaxy evolution. 
The high quality ACS imaging data combined with the FORS2 spectra allowed us to derived accurate 
galaxy photometry, and the with the additional NIR J- and Ks-bands we performed an SED analysis.

\begin{itemize}
 \item From the X-ray spectral analysis we obtained a global cluster temperature of 6.4 keV and a 
luminosity $L_{x}[0.5-2.0] keV$=1.3 $\times 10^{44}$ erg s$^{-1}$, indicating that XMM1229 is 
a massive cluster. Fixing the redshift to the spectroscopic value we obtain the metal abundance 
Z/Z$_{\sun}$ = 0.34 $^{+0.14}_{-0.13}$.
 \item We measured the cluster velocity dispersion $\sigma$ using the 27 galaxy redshifts 
obtained with FORS2, $\sigma$=683$\pm$62 km/s. The velocity dispersion is below the one expected 
from the mean T-$\sigma$ relation (\cite{rosati02}) for the cluster temperature, however it is still 
within the large scatter of the relation.
 \item Using the morphological templates of M. Postman we made a visual classification 
of the cluster galaxies. This evaluation indicates a predominance of ellipticals (15/21), with only 
four members classified as S0, one irregular galaxy and one late-type Sb galaxy. 
In order to investigate whether the shortage of S0s and also to populate the faint end 
of the cluster red-sequence, we constructed a "red-sequence" sample, based on the 
galaxies morphology, colour and total magnitude. We find that the fraction of ellipticals in the 
locus of the red-sequence pertaining to the latter galaxy sample, 22/31, is a factor three larger 
than the the number of S0s in the spectroscopic sample (9/31). 
Furthermore, there are two pairs of red galaxies in the spectroscopic sample.
 \item In addition to the visual assessment we also fitted Sersic models to the surface 
brightness profiles of the two galaxy samples.
The distribution of the best-fit structural parameters $n$ peaks at 3.9 suggesting 
a majority of bulge dominated galaxies. The median effective radius is 0.275", approximately 
the radius chosen for measuring the $i_{775}$-$z_{850}$ colour (r=0.25"). 
 \item Two methods were explored to measure the scatter of the CMR: $(i)$ in a first 
approach, as standard in the literature, we correct the different PSFs of the $i_{775}$ and $z_{850}$ 
bands to measure accurate galaxy aperture magnitudes,
and $(ii)$ in an alternative approach, we use the best-fit galaxy model magnitudes obtained by 
fitting the surface brightness profiles. 
The CMR at this high redshift is found already to be very tight, with an intrinsic scatter of 
0.04 mag when taking into account only the passive members, a spread which is similar to the 
local clusters, thus confirming that the cluster ETGs assembled early on and in short timescales.
The scatter of the red-sequence is essentially the same from these two independent methods, 
showing that the second method is robust against uncertainties arising from PSF corrections. 
The slope of the red-sequence including only the cluster members is -0.031$\pm$0.016, 
and slightly decreases to -0.022$\pm$0.008 when accounting also for the "red-sequence" 
galaxies. 

These results are in agreement with the conclusions drawn from the ACS Intermediate Redshift Cluster 
Survey (see e.g., \cite{mei06}, \cite{blake03}, \cite{mei08}), where no significant redshift evolution was found 
in the CMR scatter and slope. It is worth noting that in the referred papers, galaxy colours were measured in apertures
 of variable size corresponding to the effective radius.

 \item The spectrophotometric analysis shows a red-sequence populated by moderately massive 
galaxies, with a median stellar mass of 7.4 $\times 10^{10}$ M$_{\sun}$. 
The combined SED + spectral fit to the stacked spectrum of the passive members allowed us to 
constrain the ages of the ETGs to 3.7 $^{+0.4}_{-0.5}$ Gyr, corresponding to a formation 
redshift $z_{f}=3.0 \pm 0.5$, similarly to other z $\sim$ 1 clusters (e.g \cite{gobat})
\item The inferred star formation histories imply that the cluster galaxies have completed most of 
the chemical enrichment, which is consistent with the high metal abundance of the ICM, 
Z $\thicksim$ 1/3 Z$_{\sun}$, as derived from our X-ray analysis (see Table~\ref{xmm-results}).
\item As widely reported in the literature, we find a clear signature of significant $downsizing$, 
since the correlation between stellar mass and galaxy age favors an anti-hierarchical behavior 
where the most massive galaxies are the oldest, which also tend to be closer to the cluster core 
(Fig.~\ref{Figmassage}, Fig.~\ref{Figdaist}).

\end{itemize}

\begin{acknowledgements}
We acknowledge the excellent support provided by the staff at the Paranal observatory. 
In particular, we wish to acknowledge their assistance in setting up the observations 
with the MXU mode of FORS2 when technical problems prevented us from using the MOS mode.
We thank M. Postman for providing us with his templates for the galaxy morphological classification.
JSS would like to thank D. Pierini, M. Nonino, S. Borgani and M. Girardi for useful discussions.
JSS acknowledges support by the Deutsche Forschungsgemeinschaft under contract BO702/16-2.
RG acknowledges support by the DFG cluster of excellence “Origin and Structure of the Universe”
(www.universe-cluster.de).
This research has made use of the NASA/IPAC Extragalactic Database (NED) which is 
operated by the Jet Propulsion Laboratory, California Institute of Technology, under 
contract with the National Aeronautics and Space Administration.
\end{acknowledgements}

\end{document}